# MEMRISTIVE GaN ULTRATHIN SUSPENDED MEMBRANE ARRAY

Mircea Dragoman[1], Ion Tiginyanu[2,3], Daniela Dragoman[4,5], Tudor Braniste[3], Vladimir Ciobanu[3]

[1] National Research and Development Institute in Microtechnology, Str. Erou Iancu Nicolae 126A, Bucharest 077190, Romania

[2] Institute of Electronic Engineering and Nanotechnologies, Academy of Sciences of Moldova, Chisinau 2028, Moldova

[3] National Centers for Materials Study and Testing, Technical University of Moldova, Chisinau 2004, Moldova

[4] Univ. Bucharest, Physics Faculty, P.O. Box MG-11, Bucharest 077125, Romania

[5] Academy of Romanian Scientists, Splaiul Independentei 54, 050094, Bucharest, Romania

## Abstract

We show that ultrathin GaN membranes, with a thickness of 15 nm and planar dimensions of $12 \times 184 \ \mu m^2$, act as memristive devices. The memristive behavior is due to the migration of the negatively-charged deep traps, which form in the volume of the membrane during the fabrication process, towards the unoccupied surface states of the suspended membranes. The time constant of the migration process is of the order of tens of seconds and varies with the current or voltage sweep.

_____________________________________________________________

Corresponding author: mircea.dragoman@imt.ro

**1. Introduction**

Although the memristor was predicted long time ago, its experimental evidence occurred only when the nanotechnologies have reached a certain degree of maturity. A memristor is a unique circuit element, the fourth fundamental circuit element, which plays a similar role as the inductance (L) and capacitance (C). Unlike the inductance and capacitance, however, the memristor is a nonlinear circuit element, displaying a voltage-current dependence with a distinct footprint – a pinched hysteretic behavior of the current when the voltage is varied from negative to positive values [1]. Moreover, the resistance of the memristor depends strongly on the history of the applied voltage. Since the memristor remembers its previous state when the excitation is off, it is non-volatile and strongly related to resistive switching memories, which are considered to be the next generation of electronic memories. As such, memristors could be used in reconfigurable logic circuits [2] or in neuromorphic systems [3] in order to mimic the synapses, which are the fundamental key components of neural systems.

These and other applications have initiated a rush for inventing various memristive devices, several memristor classes being already identified. For instance, there are CMOS-compatible memristors based on (i) resistive switching memories in oxides such as $TiO_2$ and $WO_3$, (ii) electrochemical metallization (redox memristors), and (iii) phase change materials [4,5]. The memristive behavior is encountered in many materials, in which different physical effects are present. The ultimate memristor is a gate-tunable single atomic sheet of matter such as $MoS_2$, in which the memristive behavior is due to the variation of carrier density as a function of the gate voltage [6].

GaN, also termed as "new silicon", is presently the second most important semiconductor after Si, and has well-known applications in high-frequency and power electronics. In the form of nanowires and nanotubes, GaN has increasingly many applications in the area of nanoelectronics [7]. GaN memristors would further increase the applications of this material in applied physics and nanotechnologies. The imprint of memristive phenomena in GaN could be traced in effects of unipolar [8] and ambipolar resistive switching [9] in amorphous GaO, where the effect is due to the formation/rupture of conductive filaments that form due to oxygen vacancies or to the field-induced migration of oxygen vacancies, respectively, as well as in bipolar switching of the metal-insulator AlN/n-GaN structure [10] due to a trap-controlled space charge current. In the GaN membrane arrays presented in this paper, the memristive behavior is a combined mechanism, in which both trap migration and trap-controlled space charge contribute to the observed voltage and time dependencies of the current.

## 2. Fabrication of GaN membranes

The GaN memristive device reported in this paper was fabricated using a modified version of the surface charge lithography (SCL) that is well described in [11-14] and will not be repeated here in detail. We note that the SCL is based on ion-beam-writing of surface negative charge followed by photoelectrochemical (PEC) etching of the GaN layer. The ultrathin GaN membrane array was fabricated on the same MOCVD-grown GaN layer by treating some regions with 0.5-keV $Ar^+$ ions at the dose of $10^{11}$ $cm^{-2}$, as described in [11,12].

The metallic Ti/Au ohmic contacts were deposited on as grown GaN via photo-lithographically defined windows. First, 50 nm Ti was deposited using electron beam evaporation from a Ti wire, followed by thermal evaporation of 200-nm thick Au (TedPella 99.99%). The lift-off process was followed by 0.5-keV Ar$^+$ ion treatment with subsequent PEC etching of the samples, resulting in the ultrathin suspended GaN membranes. In the ion-treated regions conductive GaN membranes are formed, as shown in Fig. 1(a). In this way, we can fabricate suspended GaN membrane arrays separated by nonconductive paths. The thickness of the GaN membrane measured using SEM was found to be 15.6 nm. The SEM of the cross section of the device, shown in Fig. 1(b), reveals that the GaN membranes are sustained by a network of pillars formed by nanowires that represent threading dislocations [11]. As will become clear in the following, this way of fabricating GaN membrane arrays is at the origin of the memristive phenomena investigated in this paper.

In Ref. [14] it is shown that the diffraction pattern of GaN membranes as those used in this paper displays a wurtzite crystalline structure. A further structural characterization was performed by High-Angular Dark Field Scanning Transmission Electron Microscopy (HAADF-STEM) using a FEI Titan 80/300 instrument equipped with a corrector for spherical aberration of objective lenses and a Fishione model HADF detector. The SEM image taken from a piece of a broken membrane reveals details of the spatial architecture (see Fig. 2(a)), while the STEM image in Fig. 2(b) reveals the crystalline structure of the membrane. The thickness of the GaN membrane using HAADF-STEM techniques confirmed the nanoscale thickness of the membrane.

### 3. Measurements and discussions

The *I-V* dependence of the suspended GaN membrane array was measured using a semiconductor characterization system Keithley 4200 SCS. The DC probes connected to the Keithley 4200 were placed together with the probe station in a Faraday cage provided by Keithley and all measurements were done at room temperature. Figure 3 illustrates the experimental set-up, while the current-voltage dependence at consecutive sweeps is represented in Fig. 4. The voltage is swept first from -5 V up to +5 V and then backwards, from +5 V to -5 V. This procedure is repeated several times, the successive sweeps being marked on Fig. 4 as 1, 2, 3, and 4. It can be seen that the GaN membrane array behaves like a memristor because the *I-V* dependence in the range -5 V to + 5 V is strongly nonlinear and has a pinched hysteretic shape. Moreover, at consecutive sweeps the initial pinched hysteresis is shifted towards higher currents but preserves its shape, illustrating clearly another property of the memristor: memory, i.e. dependence on the previous state. We have investigated several devices, with membranes widths varying with few nm, as evidenced by SEM. We have performed the electrical characterization on several devices and for different voltage steps, and have not introduce any smoothing procedures intentionally, neither in the DC measurement station computer nor in the computed results after measurements. We have not observed any significant changes in the current-voltage dependencies in different experimental conditions.

Similar increases of the current at successive voltage sweeps have been observed in other materials as well, and was explained by a trap-controlled space charge limited current mechanism [15], in which negatively-charged deep traps generate electric fields with the same orientation as the applied fields. In order to investigate the possibility that charge transport in our device is determined by the same mechanism, we have represented in Fig. 5 the dependence $\log(I)$ versus $\log(V)$ for the first and the fourth sweep in Fig. 4, looking for the value of $\alpha$ in the relation $I = V^\alpha$. As can be seen from Fig. 5, the $\alpha$ values are practically the same for the first and the fourth sweep. For low applied voltages $\alpha \cong 1$, which corresponds to an Ohmic behavior, while for higher voltages the nonlinearity increases, the values of $\alpha$ increasing continuously up to about 4. These values are consistent with a space charge limited current mechanism [10,15], in which the electric field induced by the negatively-charged deep traps enhances the applied field. In the case of the GaN membrane array, the negatively-charged deep traps are formed during irradiation by Ga$^+$ ions.

Figure 6 represents the dependence of the current on time at the same applied voltage of –7 V, maintained constant during the experiment, the data being collected during 3 minutes for five consecutive time sweeps. From Fig. 6 it can be seen that at each sweep the amplitude of the current increases, although the applied voltage remains the same. The respective voltage-time dependence at constant current, illustrated in Fig. 7, shows a decrease of the voltage in time. The compatible time dependences of both current and voltage can be explained by an electric-field induced migration of trapped negative charges towards the surface states of the membrane. The migration is only observed for high-enough applied voltages or currents, for which the barriers confining the trapped charges decrease; we have not observed a consistent time variation of currents and/or voltages for smaller values of voltages/currents than those in Figs. 6 and 7, which suggests that the time dependences are not a result of a simple detrapping process. The time constants associated to the migration of trapped charges are different for different sweeps. For instance, for the first to the fifth sweeps in Fig. 6 they are: 9.3 s, 18.2 s, 17.5 s, 15.8 s, and 11.2 s, respectively, while the corresponding migration times for the first and the second sweep in Fig. 7 are 40 s and 69 s. The increase of the migration time from the first to the second sweep reflects the migration to the surface states of charges on deep traps situated increasingly far from the surface, whereas the slight decrease of this parameter for further sweeps could be attributed to the build up of an electric field close to the edges of the membrane that could accelerate the migration of the remaining trapped charges.

So, due to the increasing applied electric field, the charges trapped in the membrane volume migrate towards the membrane surface and fill the surface states, thus increasing the current. The surface states are filled gradually, the current increasing progressively at each constant-voltage cycle. The minimum resistance value, denoted by $R_{ON}$, corresponds to the situation when all surface states are filled due to charge migration whereas the maximum value of this parameter, $R_{OFF}$, is associated to the case when no surface state is filled, i.e. the device is in the initial state. The charge migration in our GaN membranes is analogues to the drift of oxygen vacancies, seen as positively-charged dopants, in the first memristor, based on a $TiO_2$ oxide layer with a thickness of 5 nm sandwiched between two Pt contacts [16]. In the latter case, $R_{ON}$ and $R_{OFF}$ are attributed to doped and undoped states of the $TiO_2$. Based on this analogy, the time dependence of the current-voltage characteristics can be described as [3,16]:

$$i(t) = \frac{v(t)}{R_{ON}\gamma(t) + R_{OFF}[1 - \gamma(t)]} \qquad (1)$$

where $\gamma(t)$ is a continuous time function with values in the [0,1] domain. The function $\gamma(t)$ reaches its maximum and minimum values, of 1 and 0, when $R = R_{ON}$ and, respectively, $R = R_{OFF}$. $\gamma(t)$ is considered as a linear function of the flux linkage $\Phi(t) = \int v(t)dt$ [3], so that the memristor equation as defined initially by Chua is retrieved. Equation (1) determines the equivalent circuit of GaN membrane memristive device as being formed from two variable resistances connected in series, like in [16]. In the case of our membranes, $R_{ON} = 806\ \Omega$ and $R_{OFF} = 77\ k\Omega$, the ratio $R_{OFF}\,/\,R_{ON}$ attaining the value of 95.

Regarding the power supported by the GaN membrane, the maximum power estimated from the data in Fig. 4 is about 6 mW at 5 V in the first sweep and 10 mW in the fourth sweep. However, we have experimentally determined that a single GaN membrane can support currents as high as 60 mA at 9 V, and thus a power of 540 mW, no signs of failure being observed during the measurements. The reason is that the GaN wurtzite crystal combines two essential physical properties for high-power applications: a large bandgap, of 3.4 eV, and a high thermal conductivity, of 1.3 W/(cm·K) at room temperature. Therefore, besides the many high-power circuits and integrated circuits currently based on GaN, the GaN membrane memristor reported in this paper is a high-power device compared to other memristors reported in the literature. Because the metallic contacts with the sample are ohmic and the membrane is supported on isolator pillars, the electrical losses in the device are negligible.

## 4. Conclusions

In conclusion, the GaN membrane array behaves like a memristor as a result of the fabrication method, which on one side produces defects in the GaN membrane that trap negative charges, and on the other side creates a very large amount of unoccupied surface states, much larger than the number of defects. Under the action of the electric field, the trapped charges in membranes migrate by hopping or tunneling from one defect to another until they reach the surface states. The surface states are increasingly filled with electrons at every voltage sweeps, the consequence being an increase in the current, as shown in Fig. 4, as a result of a decreased screening of the applied electric field. It is clear from Figs. 6 and 7 that the memristor memorizes its previous state, but at each sweep the current and the voltage vary in time due to the fact that the number of trapped negative charges in the volume decreases as a result of their migration to the surface states. The time evolution of the current and voltage takes place until all volume trapped charges migrate towards the surface. This delicate balance between trapping and hopping/tunneling, defects and surface states, is at the origin of the behavior of the GaN membrane array memristor.

*Acknowledgements*

We would like to thank to Thorsten Mehrtens and Andreas Rosenauer from Institute of Solid State Physics, University of Bremen, Germany for HAADF-STEM measurements.

M.D. would like to dedicate this paper to the 150 years anniversary of Romanian Academy and to the memory of our Humboldt colleagues, professors Radu Grigorovici and Rodica Manaila, members of the Romanian Academy.

**Figure captions**

Fig. 1   SEM images of (a) the GaN membrane array and (b) of the cross-section of the GaN membrane device for membrane thickness measurements.

Fig. 2 (a) SEM image of broken GaN membranes, and (b) STEM image of the broken GaN membrane.

Fig. 3   The DC experimental set-up for GaN membrane measurements

Fig. 4   Current-voltage dependence at consecutive sweeps.

Fig. 5   Log($I$) versus log($V$) dependence of the first and fourth sweep in Fig. 2.

Fig. 6   Current-time dependence at constant voltage and consecutive sweeps.

Fig. 7   Voltage-time dependence at constant current and consecutive sweeps.

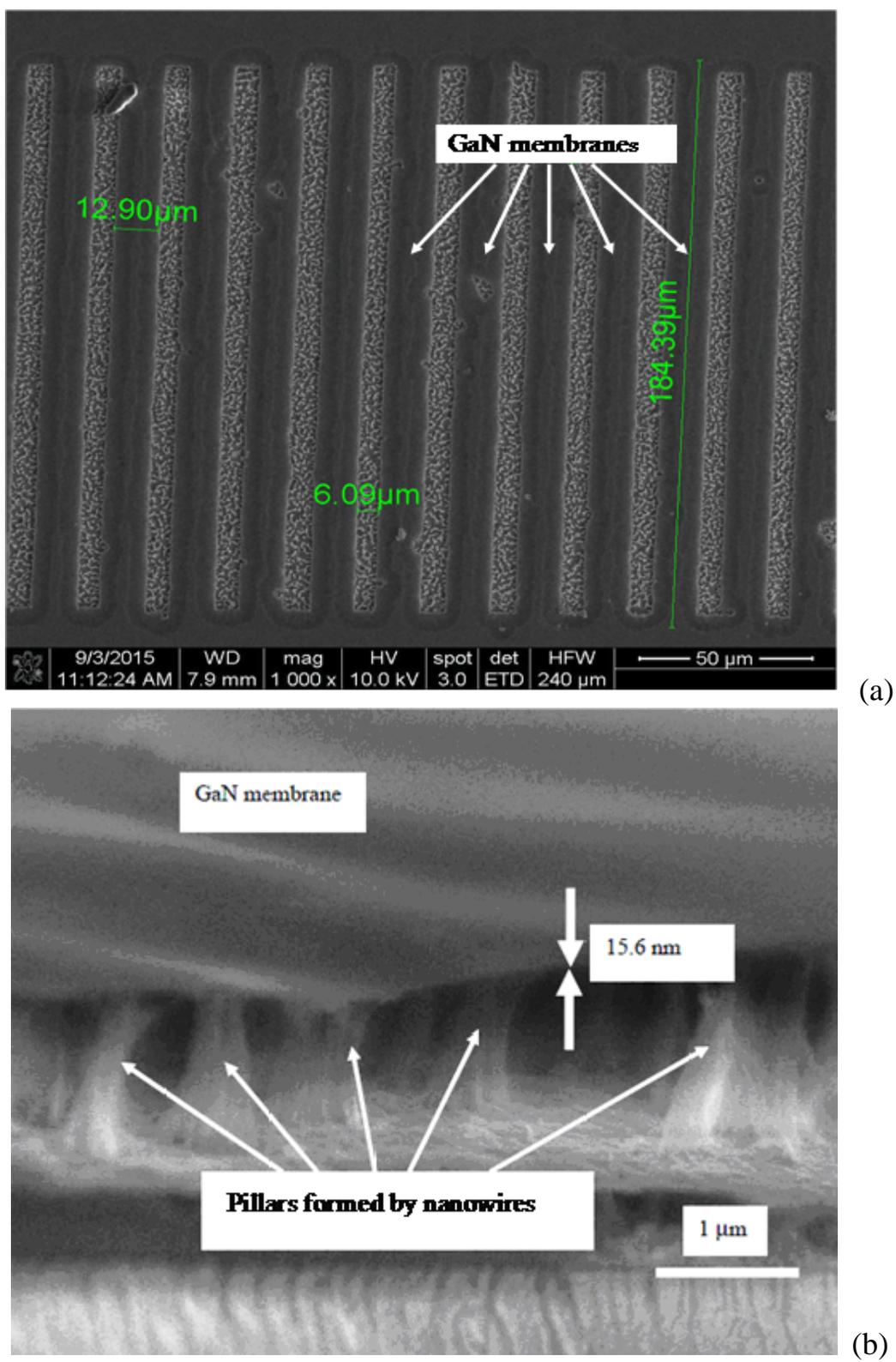

(a)

(b)

Fig. 1

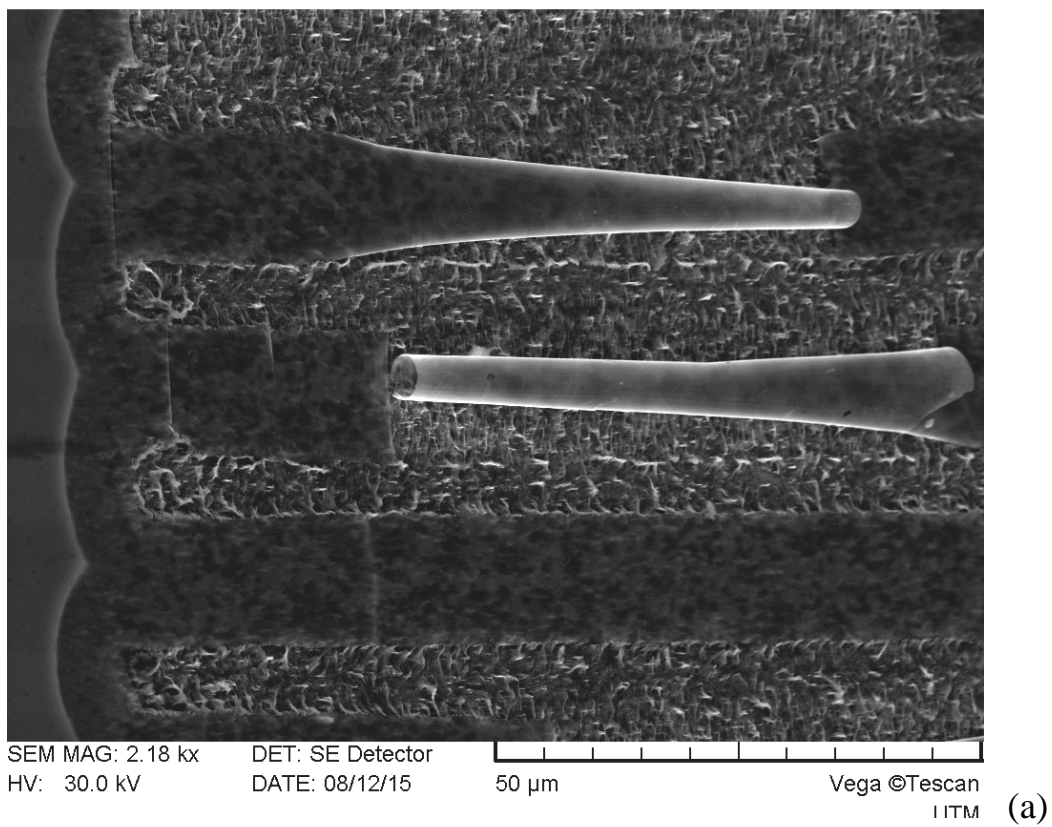

SEM MAG: 2.18 kx    DET: SE Detector
HV:   30.0 kV       DATE: 08/12/15    50 µm                Vega ©Tescan
                                                           UTM   (a)

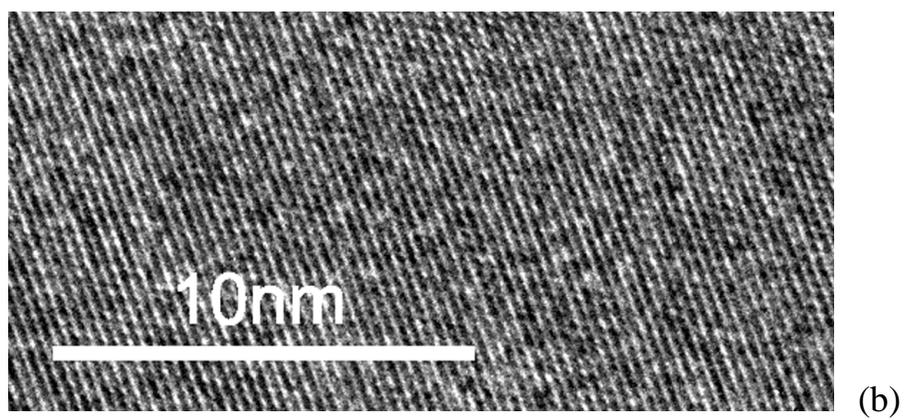

10nm

(b)

Fig. 2

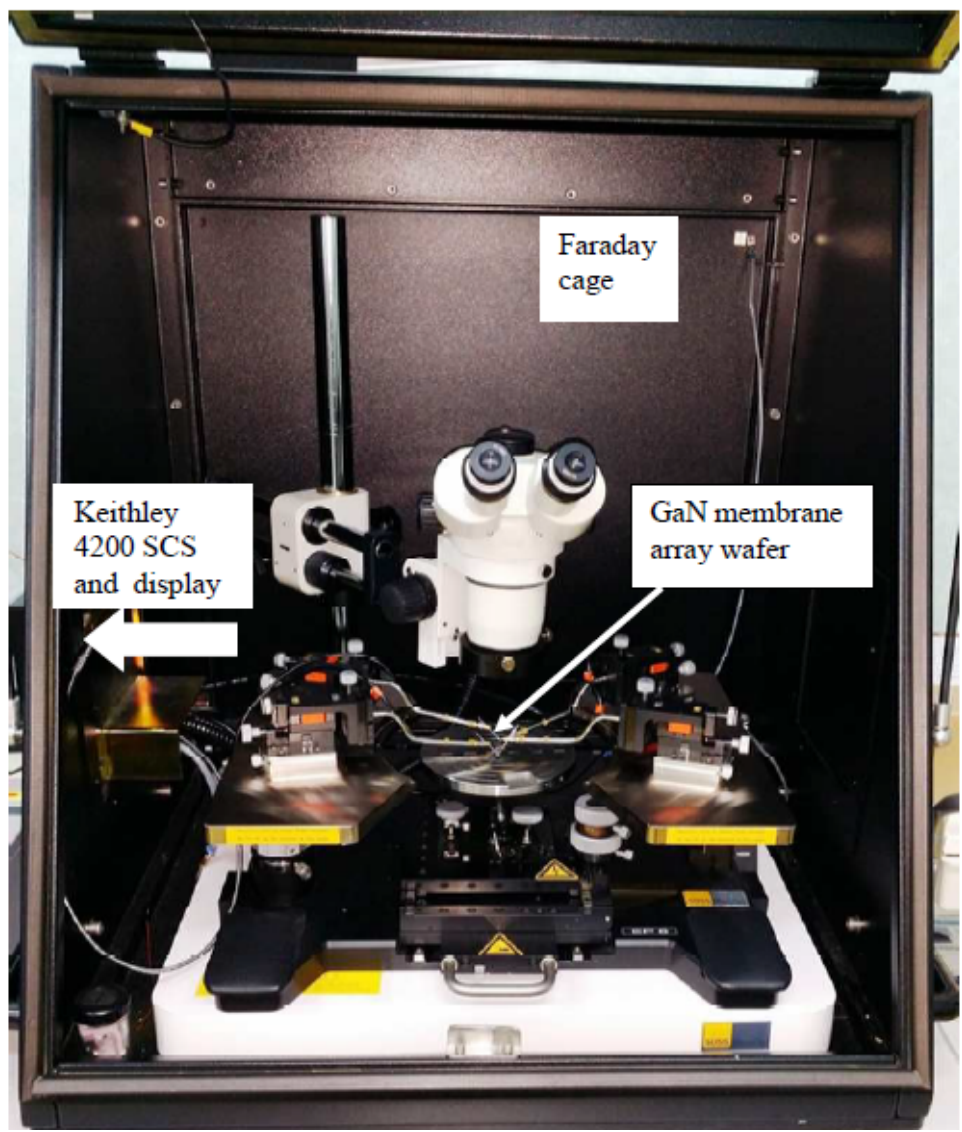

Fig. 3

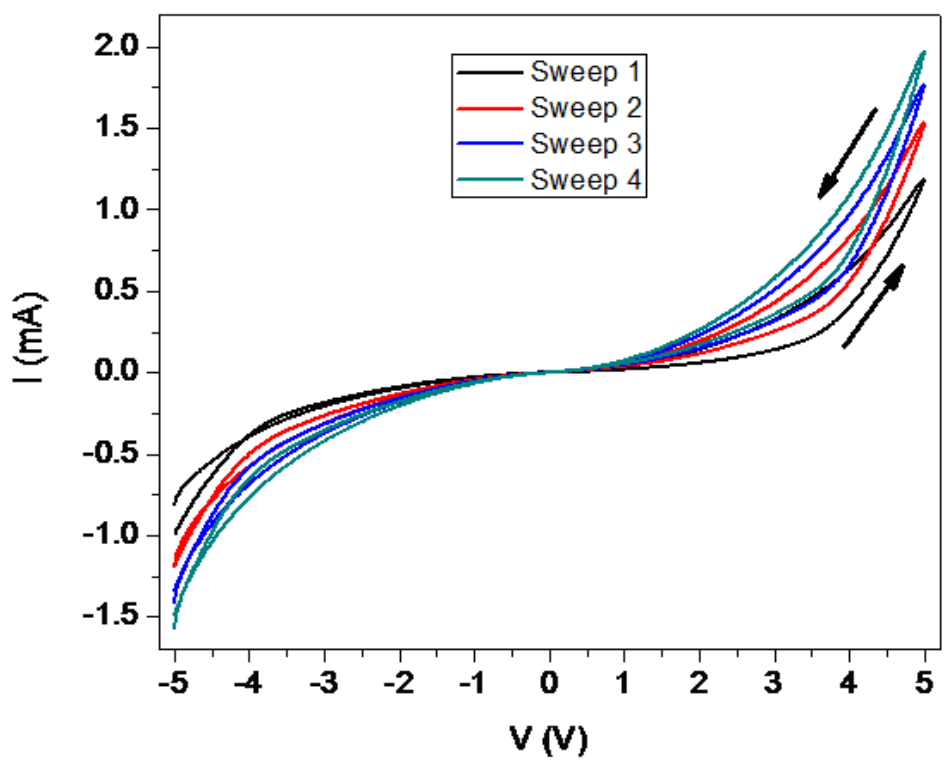

Fig. 4

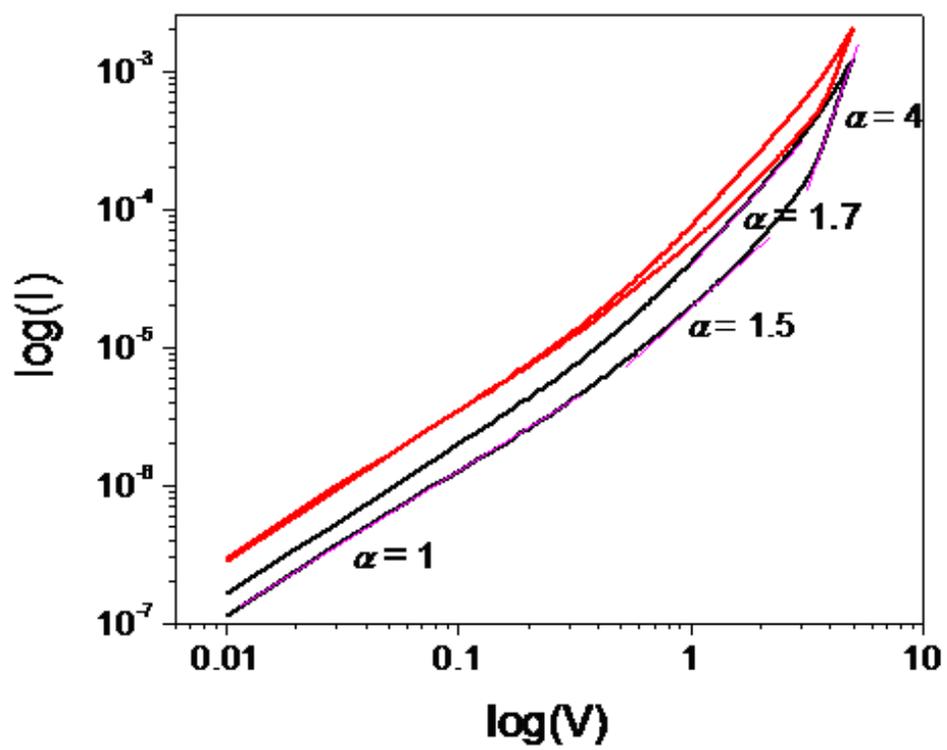

Fig. 5

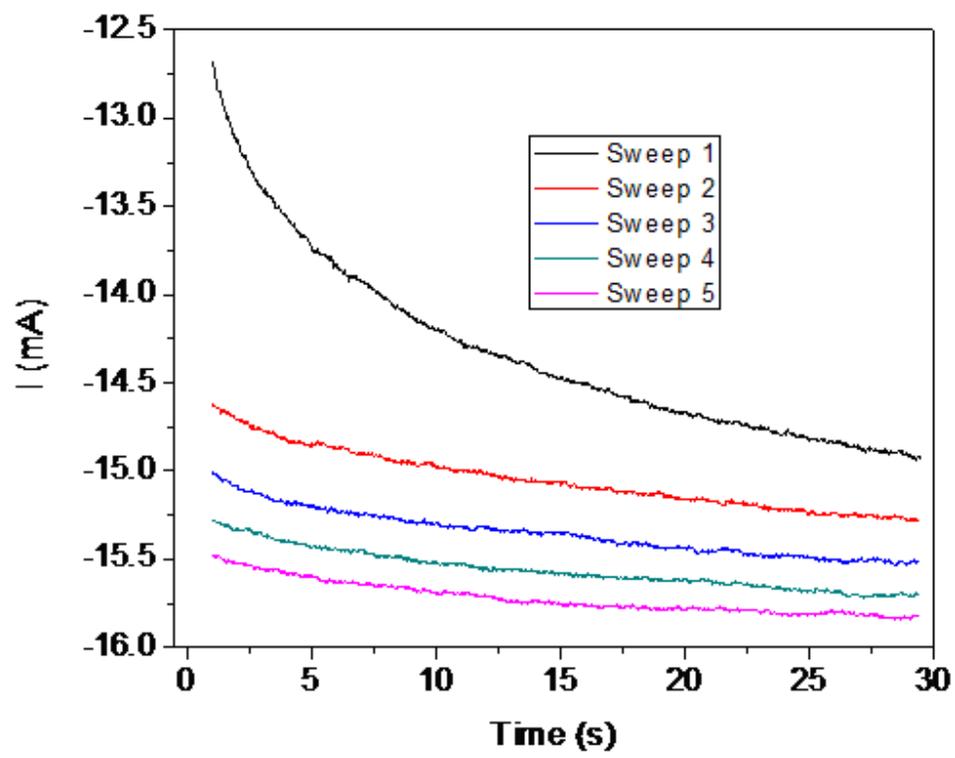

Fig. 6

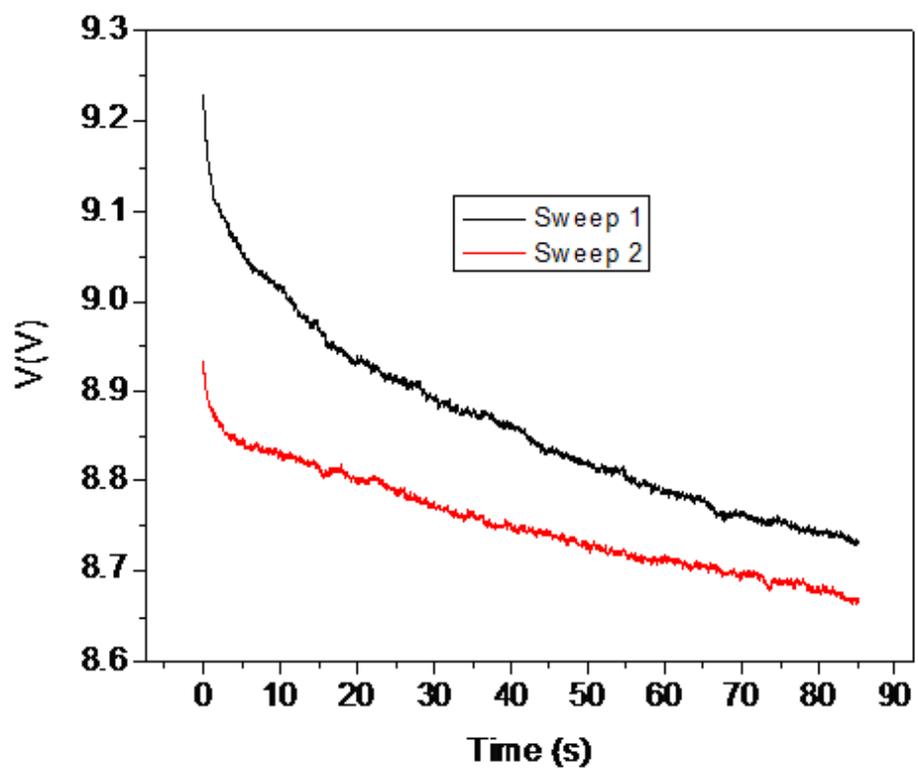

Fig. 7